\documentclass[journal=jctcce,manuscript=article]{achemso}

\usepackage[version=3]{mhchem} 
\usepackage{algorithm}
\usepackage{algorithmic}
\usepackage[hidelinks]{hyperref}
\usepackage{float}
\usepackage{graphicx}
\usepackage[user,titleref]{zref}
\UseRawInputEncoding

\author{Hassan Nadeem}
\author{Diwakar Shukla}
\email{diwakar@illinois.edu}
\affiliation[BioE]
{Department of Bioengineering, University of Illinois at Urbana-Champaign, Urbana, IL, 61801, USA}
\alsoaffiliation[CHBE]
{Department of Chemical and Biomolecular Engineering, University of Illinois at Urbana-Champaign, Urbana, IL, 61801, USA}
\alsoaffiliation[Chem]
{Department of Chemistry, University of Illinois at Urbana-Champaign, Urbana, IL, 61801, USA}
\alsoaffiliation[Biophysics]
{Center for Biophysics and Quantitative Biology, University of Illinois at Urbana-Champaign, Urbana, IL, 61801, USA}
\title[adaptive sampling]
  {Optimizing adaptive sampling via Policy Ranking}

\keywords{Molecular Dynamics, Adaptive Sampling, Conformational sampling, Ensemble sampling}

\begin{document}

\begin{abstract}
Efficient sampling in biomolecular simulations is critical for accurately capturing the complex dynamical behaviors of biological systems.  Adaptive sampling techniques aim to improve efficiency by focusing computational resources on the most relevant regions of phase space. In this work, we present a framework for identifying the optimal sampling policy through metric driven ranking. Our approach systematically evaluates the policy ensemble and ranks the policies based on their ability to explore the conformational space effectively. Through a series of biomolecular simulation case studies, we demonstrate that choice of a different adaptive sampling policy at each round significantly outperforms single policy sampling, leading to faster convergence and improved sampling performance. This approach takes an ensemble of adaptive sampling policies and identifies the optimal policy for the next round based on current data. Beyond presenting this ensemble view of adaptive sampling, we also propose two sampling algorithms that approximate this ranking framework on the fly. The modularity of this framework allows incorporation of any adaptive sampling policy making it versatile and suitable as a comprehensive adaptive sampling scheme.
\end{abstract}

\section{Introduction}

Molecular dynamics (MD) simulations have emerged as indispensable tool to uncover dynamics and evolution of various physical systems at atomic scale. Through numerical solutions of Newton's equations of motion, MD simulations provide a microscopic scale view of the time evolution of these systems. For biomolecular systems, MD simulations have shed light on complex processes such as protein folding \cite{LindorffLarsen2011} , enzyme activity\cite{Romero2019} , transport activity \cite{KhaliliAraghi2009, Chan2021} and cellular signaling \cite{Latorraca2016, Dutta2022}. In addition to providing a fine-grained description of the underlying process, these simulations allow for a deeper understanding of complex relationships between structure, dynamics and functions of these systems. 

Despite presenting a detailed atomic level description that is unparalleled; MD simulations suffer from the following limitations. The output trajectories from MD simulations are high-dimensional, difficult to interpret and require significant post-processing to provide an elegant low-level description of the underlying processes. In addition the accuracy of the MD simulations is heavily tied to the accuracy of the interaction model that was used to parameterize the governing equations. While several such models exist (CHARMM\cite{MacKerell1998} , AMBER \cite{Tian2019} , OPLS\cite{Robertson2015}, GROMOS\cite{Oostenbrink2004}  etc.), a generalizable model that replicates physical systems to an arbitrary accuracy across various materials and bio-molecular systems, still eludes us. In order to maintain the fine-grained level of detail that MD simulations promise, the integration time-step is constrained by the higadaptivh-frequency bond vibrations. Even with modern bond restraining algorithms\cite{Ryckaert1977,Andersen1983,Miyamoto1992} and mass-distribution schemes\cite{Mao1991,Feenstra1999}, this limiting factor only allows the time-step range to be between 1-4 fs. Most physiologically relevant phenomenon like changes in protein conformation, ligand binding to a G-Protein coupled receptor or sugar transport through a membrane-transporter, all occur at time-scales ranging from microseconds to milliseconds\cite{LindorffLarsen2011,Kubelka2004}. This presents a computational challenge of performing billions of time-step computations to simulate these processes. Even with modern GPUs and rapidly advancing computational resources, this makes these simulations intractable for most, but specialized super-computing facilities. Lastly these simulations have the propensity to remain stuck in meta-stable states. These states may be unproductive intermediate states that the simulation exhausts a lot of resources on.

To address the last two limitations several methodologies have been proposed. Enhanced sampling\cite{Henin2022,Kleiman2023} algorithms present a popular choice. These methods aim to improve the efficiency of sampling by accelerating the exploration of the system's configurational space, thus enabling the study of processes that occur on longer timescales or in less accessible regions of the energy landscape.   Adaptively biased simulations \cite{Laio2002,Barducci2008,Valsson2016, Valsson2020}, e.g. Metadynamics , are such techniques in which an external bias is added to the system, which is learned from progression of the simulation trajectory. In Metadynamics, repulsive Gaussian kernels are added in the collective variable space to lift sampling by giving it an extra `push' towards the rare events. For such techniques free energy surfaces can be recovered via reweighing, but recovering kinetics is non-trivial unless simplifications\cite{Donati2017} are assumed. Coarse-graining (CG)\cite{Souza2021,Moore2014,Wang2009} is another method which attempts to accelerate sampling by addressing the time-step limitation. In CG approaches, atoms are lumped into chemically representative beads which are then dynamically simulated. This lumping reduces the number of particles, thereby reducing the number of computations required per time-step. In addition this approach gets rid off the high-frequency hydrogen bonds which allows a higher time-step of 10-20fs. This simplification is achieved with the loss of chemical detail which is the driving force of many intra and intermolecular interactions.  

In contrast to adding external bias or coarse-graining, adaptive sampling\cite{Huang2009,Bowman2010, Kleiman2023} is a technique which aims to sample the configuration space by prioritizing sampling from regions which enable efficient exploration of the ensemble. Complex biomolecular processes such as protein folding are marked by rare transitions along a pathway to the natively folded state. The probability of overcoming a free energy barrier to observe the rare event of protein folding, depends upon the number of attempts made to overcome this barrier. Hence, the cumulative simulation time and not the length of the single long simulation is significant in this case \cite{Zimmerman2018,PhysRevLett.86.4983}. Adaptive sampling exploits this by seeding simulations from strategically chosen configurations which enable rapid exploration of the configuration space. Figure 1(A) shows the typical flow of an adaptive sampling strategy. At each round, parallel, short simulations are run to perform sampling. The resultant trajectories are featurized into a lower-dimensional representation. These features could be physically inspired choices such as residue-residue contacts or data driven choices such as  time-lagged independent components\cite{Molgedey1994,No2015} (tICs). These simplified trajectories are then discretized by traditional clustering algorithms e.g. KMeans\cite{1056489} into (micro)states. At this stage states are selected to start subsequent round trajectories according to the choice of the adaptive sampling scheme. Typically, no external bias is added to the system, therefore the selection strategy is the only distinguishing choice in the adaptive sampling policies. Various adaptive sampling policies have been proposed for this purpose.

\begin{figure}[!htb]
    \centering
    \includegraphics[width=\textwidth]{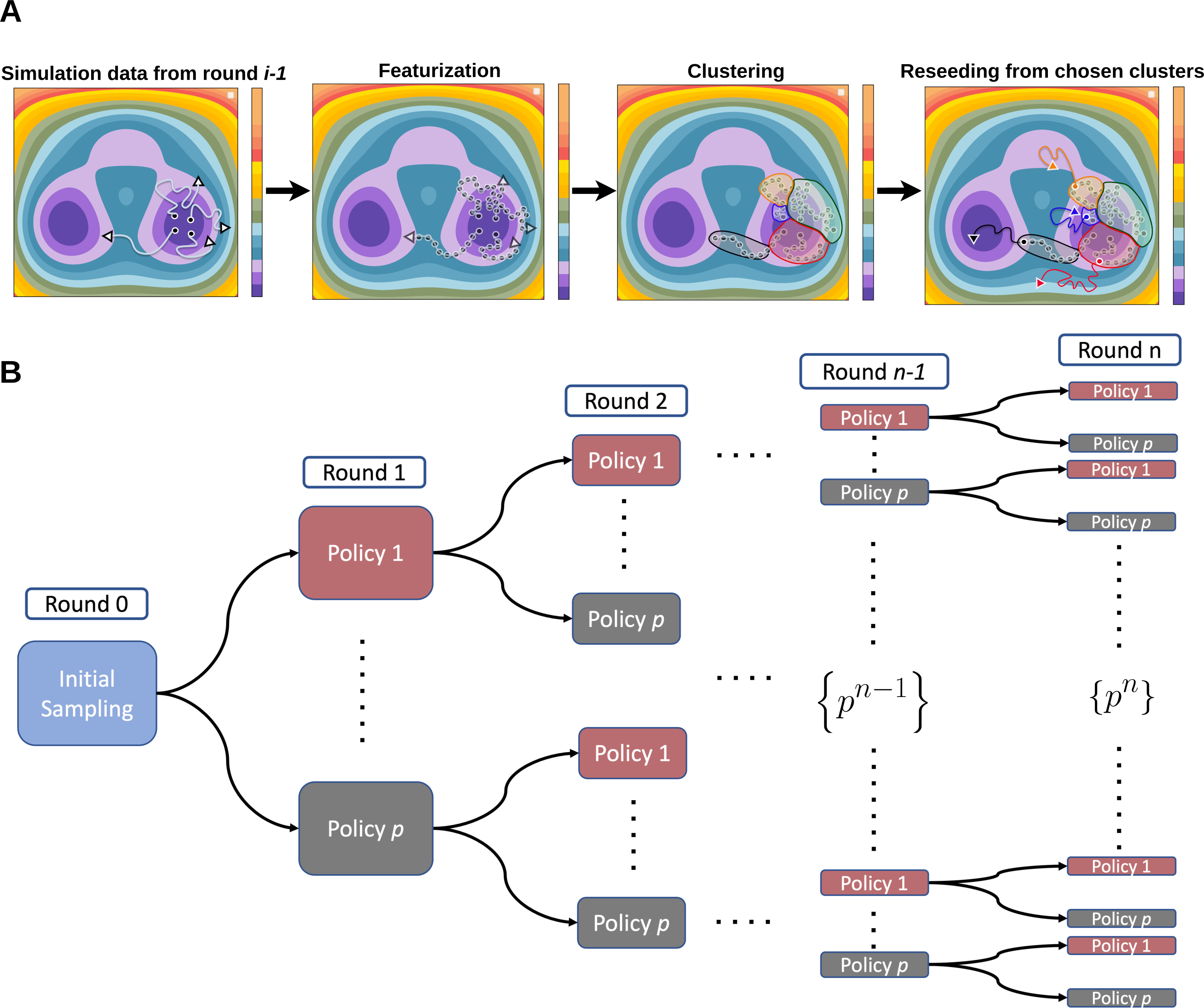}
    \caption{(A) A typical adaptive sampling workflow. Short parallel simulations are run and then featurized. Feature space is then discretized into microstates by clustering and reseeding is performed according to the chosen adaptive sampling policy.(B) Evolution of the policy space with $p$ policies. The number of paths in the tree increase exponentially as $p^{n}$, where $n$ is the number of adaptive sampling rounds.}
    \label{fig:figure_1}
\end{figure}

The first class of adaptive sampling policies can be described as Markov State Model (MSM) inspired or Count based strategies. The simplest amongst these is Random Sampling, where states are randomly selected to seed the next round trajectories. Least counts sampling \cite{Bowman2010} focuses on improving exploration by selecting states with the lest number of counts, which equates to prioritizing states which have been least visited in the last round of simulations. A similar scheme that shares this idea is Adjacency sampling\cite{Weber2011}. In this method the restart criterion is connectivity based, according to which states which have the least number of connections with adjacent states are selected.  Another method, termed Lambda sampling\cite{Hinrichs2007} here, was derived from the spectral uncertainty analysis of MSMs. The authors derived closed-form expressions for the distribution of eigenvalues of the transition matrix $T_{ij}$, and states which contribute the most to the variance of the first non-trivial eigenvalue were selected. The idea here is that starting sampling from these states would contribute most to the convergence of the MSM.   

Machine learning inspired adaptive sampling methods focus on efficient sampling through reward-based frameworks. Fluctuation amplification of specific traits (FAST)\cite{Zimmerman2015} was one of the first such schemes to address the exploration/exploitation dilemma of adaptive sampling. The reward function in FAST, through adjustment of a control parameter, focuses for sampling along a gradient of interest (e.g. solvent exposed surface area for protein folding), while also allowing for exploration of less sampled states. Reinforcement learning based adaptive sampling (REAP)\cite{Shamsi2018}, aims to dynamically weigh the collective variables (CVs) for the system being sampled in order to identify the least visited states. This gives an advantage where CVs are not evident from physical considerations or intuition. This idea was extended through Multi-agent (MA) REAP\cite{Kleiman2022}, in which multiple agents share a stake in reseeding and share information during sampling. Another method, Maximum-entropy VAMPNet\cite{Kleiman20233}, utilizes VAMPNets to identify states with the highest Shannon entropy for reseeding. VAMPNets, through application of a Softmax layer, can output probabilities of state assignment for each conformation, which can subsequently be scored via Shannon entropy. Other methods have also leveraged deep learning frameworks to improve conformational exploration. DeepDriveMD\cite{8945122, Gupta2024} is a framework that combines deep learning with molecular dynamics simulations to enhance the analysis of protein folding. By leveraging deep learning to derive latent representations aligned with relevant reaction coordinates, it effectively directs simulations to investigate under-sampled conformational states. This innovative method has shown substantial improvements in efficiency, achieving performance gains of up to 2.3 times in sampling folded states compared to conventional molecular dynamics approaches. Targeted Adversarial Learning Optimized Sampling (TALOS)\cite{Zhang2019} is a technique modifies the potential energy surface to encourage transitions towards a specified target distribution, thereby lowering free-energy barriers. TALOS integrates concepts from statistical mechanics with deep learning, leveraging a competitive dynamic between a sampling engine and a virtual discriminator to construct bias potentials without supervision. Tangent Space Least Adaptive Clustering (TSLC)\cite{buenfil2021tangent}, captures global manifold geometry using only local information. It replaces k-means with a custom clustering approach, CLUST, to better suit the sampling problem. Similar to REAP\cite{Shamsi2018}, it involves timestepping, clustering, reweighting, and reinitialization. Known collective variables (CVs) are weighted at each iteration to prioritize under-sampled regions, assuming the potential energy surface lies near a low-dimensional manifold. Borrowing the multi-armed bandit framework for the exploration-exploitation dilemma, AdaptiveBandit\cite{Prez2020} applies the multi-armed bandit framework to optimize adaptive sampling in molecular simulations. In this approach, each sampled conformation is treated as an arm, with the UCB1 algorithm managing the trade-off between exploring new conformations and refining promising ones. Markov State Models estimate the free energies of these conformations, which guide reward calculations. The method dynamically updates rewards and refines the action space throughout the simulation, ensuring efficient exploration while avoiding redundant sampling.

Another class of methods stems from the weighted ensemble framework. The Weighted Ensemble (WE)\cite{Zuckerman2017, Saglam2015} method is an adaptive sampling strategy designed to efficiently explore rare events in complex systems. It works by splitting the simulation into multiple trajectories, each assigned a statistical weight. These trajectories are periodically replicated or pruned, with weights adjusted accordingly, to maintain balanced sampling across different regions of the system's state space. By focusing computational effort on under-sampled pathways, WE improves the estimation of transition rates and equilibrium properties, making it well-suited for studying processes with long timescales. Beyond this broad categorization, there are several other\cite{Doerr2014,Shamsi2017,Prez2020,Blumer2022,Blumer2024,Faradjian2004,Cerjan1981,Huber1996} adaptive schemes and readers are directed towards comprehensive reviews\cite{Henin2022,Kleiman2023} for a more detailed description of these methods. 

A typical workflow of adaptive sampling simulations will involve picking a single adaptive sampling policy, based on experience, intuition or preference, and then performing each round of sampling according to the same criterion. Given an ensemble of policies $\mathcal{P}$ with cardinality $p$, Figure 1(B) shows all the possible paths that can be traversed in policy space. It is evident that the space increases exponentially with number of adaptive sampling rounds. With a choice of only two policies and 20 rounds of sampling, the total number of paths (policy choices) exceed a million choices. In general, with $p$ policies and $n$ sampling rounds, the number of paths equate to $p^{n}$. Considering the same policy is chosen at each round, the number of single policy paths is a tiny fraction i.e. $p^{1-n}$ of the total, while all the mixed paths are left unexplored. With this picture in mind we aim to address the following questions in this study. Does there exist a set of policy choices that gives \textit{optimal} sampling? In simpler terms, does a different choice of adaptive sampling policy at each round provide an advantage over single policy choice? And if so, can this knowledge be leveraged to identify this set of policy sequences on the fly?
Our results demonstrate a clear advantage of selecting a different policy at each round over single policy choice. We introduce the policy ranking framework, that not only identifies this set of policy choices but also illustrates the potential for significant improvement in terms of both exploration of conformational landscapes and statistical convergence. 

\section{Methods}\label{sec:methods}

\subsection{Markov State Models}
Markov state models (MSMs) are key components in several modules of our algorithm so we outline some foundational details of these models.
 MSMs are statistical models that describe the dynamics of a system in terms of a discrete set of states and the probabilities of transitioning between these states over time. The fundamental assumption is that the system's future state depends only on its current state (Markov property).
The Markovian property in the context of MSMs for molecular dynamics simulations asserts that the future state of a system depends only on its current state and not on the sequence of states that preceded it. Mathematically, this is expressed as:

\[
P(s_{t+1} = j \mid s_t = i, s_{t-1} = i_{t-1}, \dots, s_0 = i_0) = P(s_{t+1} = j \mid s_t = i)
\]

Here $s_t$ represents the state of the system at time $t$, $i, j$ are specific states in the system. And $P(s_{t+1} = j \mid s_t = i)$ is the probability of transitioning from state $i$ at time $t$ to state $j$ at time $t+1$.This equation indicates that the transition probability from state $i$ to state $j$ depends only on the current state $i$ and is independent of any prior states.
For the complete system dynamics, the Markov property implies that the evolution of the probability distribution $\mathbf{p}(t)$ over all states can be described by:

\[
\mathbf{p}(t+1) = \mathbf{p}(t) \mathbf{T}
\]

where $\mathbf{p}(t)$ is the row vector of probabilities of being in each state at time $t$. And $\mathbf{T}$ is the transition probability matrix where each element $T_{ij} = P(s_{t+1} = j \mid s_t = i)$.

The eigenvalues of $T_{ij}$ are fundamental to characterizing the system's kinetic behavior.The dominant eigenvalue, which is always 1 ($\lambda_{1} =1$), corresponds to the stationary distribution, representing the equilibrium state of the system.
 The subdominant eigenvalues are $\left|\lambda_{i}\right| \le 1 ; i = 2,3,4, ..$ and correspond to slower relaxation modes, with each eigenvalue associated with a distinct kinetic timescale. These timescales reflect the rates at which the system transitions between different metastable states. Eigenvalues near 1 indicate particularly slow processes, revealing critical information about the system's long-timescale dynamics, such as the stability of metastable states and the kinetics of interconversion between them.

\subsection{Ranking Policies to Achieve Optimal Sampling}

In order to investigate whether choosing a different adaptive sampling policy at each round gives optimal sampling compared to a single policy, we follow the strategy illustrated in Figure 2. Given that we have simulation data from previous round, we featurize it and apply a clustering scheme, such as KMeans clustering to discretize the featurized trajectories. After the clustering step the trajectories are passed to the \texttt{Policy Module.} The policy module selects seeds according to different adaptive sampling policies and simulations are performed for these seeds. The simulation trajectories are then passed on to the \texttt{Evaluation Module} that evaluates the sampling from each policy according to pre-defined metrics and performs a ranking. The data from the highest ranked policy is concatenated to the previous data and the process is repeated. The following sections elaborate the modules in this workflow in detail.

\begin{figure}[!htb]
    \centering
    \includegraphics[width=\textwidth]{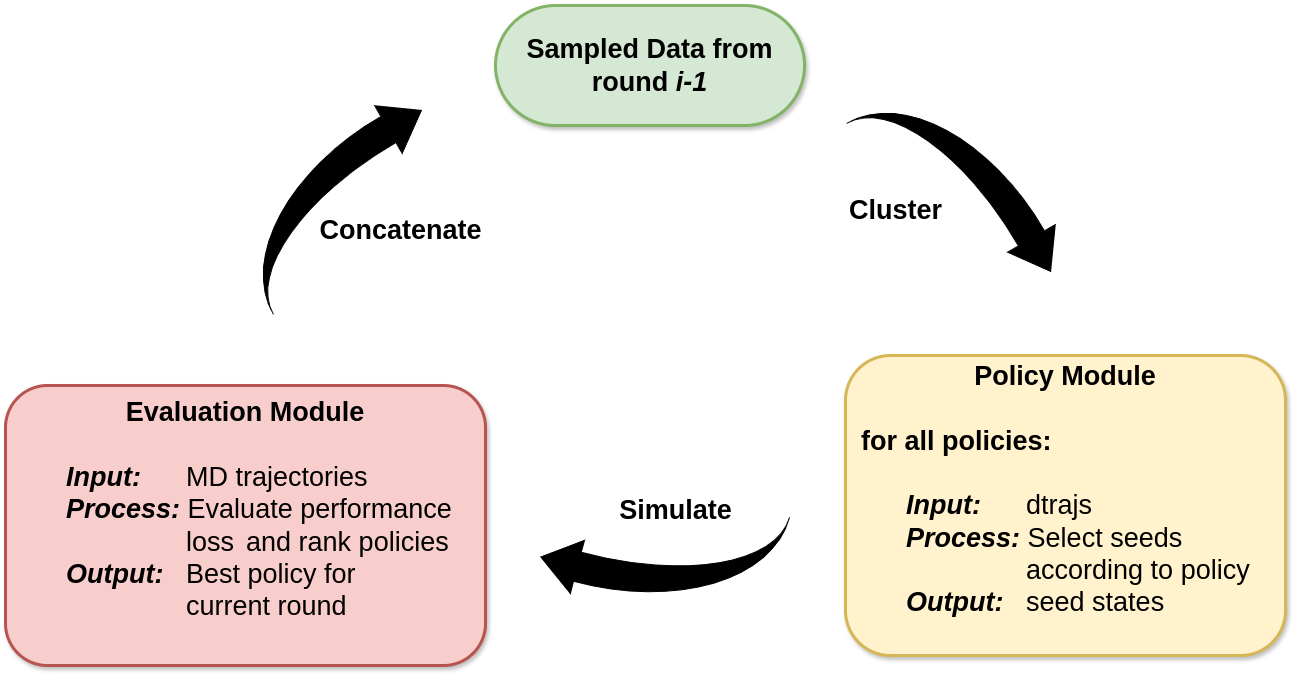}
    \caption{Simplified workflow to rank policies at each round of adaptive sampling. Data is passed to the \texttt{policy module} to produce seeds for next round. The simulation from these seeds are passed on to the \texttt{evaluation module} which ranks the policies and sampled data from the best policy is appended to the previous data.}
    \label{fig:figure_2}
\end{figure}

\subsubsection{Policy Module}
Adaptive sampling policies are designed to strategically select states from the clustered space as seeds for subsequent rounds. We define our policy ensemble and  as:
\begin{equation}
    \mathcal{P} = \{\text{Random Sampling, Least Counts, Lambda Sampling}\}
    \label{eq1}
\end{equation}
Random sampling as the name suggests chooses random states as seeds for the next round. This serves as a baseline policy as no \textit{intelligent} selection is performed and seeds are chosen at random.
Least Counts refers to selecting the least populated states. When simulation trajectories are clustered, the cluster/state indices can be sorted in ascending order by number of counts and top least visited states can be selected as seeds. The idea behind Least Counts is to select, for instance, protein conformations that have been sampled the least and therefore favour exploration into unsampled areas of the configuration space. If the simulation is stuck in an intermediate (metastable) state, Least Counts will select states at the edge of the sampled space that are least visited and give a higher chance of leaving the metastable state. Lambda Sampling refers to the adaptive sampling scheme presented by Hinrichs et al.(2007)\cite{Hinrichs2007} The framework for this policy is derived from calculations of the uncertainties in the eigenvalues and eigenvectors $T_{ij}$. The closed-form expression for the distributions of the eigenvalue $\lambda$ can be decomposed to calculate the contribution to variance from the elements in each row of $T_{ij}$. Consequently those states which have the highest contribution to variance can be chosen as seeds to increase precision of this distribution. Refer to SI methods for the mathematical expression.

The policy module processes the discretized trajectories to identify seeds for next round according to each policy in (\ref{eq1}). Simulations are then carried out using these seeds and resultant trajectories from each policy are then forwarded to the \texttt{Evaluation Module}.

\subsubsection{Evaluation Module}

The purpose of this module is to identify the \textit{optimal} policy by evaluating the performance on the current round of simulations. In order to perform a quantitative comparison, a converged MSM for each system (see Results) was constructed to act as a ground truth prior to implementing this algorithm. The MSM object and the state-definition post-clustering were saved for use in the evaluation module.
The set of trajectories for each policy are evaluated to analyse their performance in terms of exploration of the landscape as well as degree of the convergence compared to the ground truth MSM.

We defined exploration as fraction of the visited states to the total number of discretized states in definition of the ground truth MSM.
\begin{equation}
    \mathcal{E} = \frac{S_{v}}{S_{T}}
    \label{eq2}
\end{equation}

Here $S_{v}$ is the number of unique visited stated and ${S_{T}}$ is the total number of states in the ground truth MSM. A smaller value of $\mathcal{E}$ would indicate better exploration by the adaptive sampling policy.

To quantify convergence we use the relative entropy metric presented by Bowman \textit{et. al}\cite{Bowman2010} to compare MSMs. In general the relative entropy for two normalized distributions P and Q is:
\[
D(P\vert \vert Q) = \sum \limits_{i} P_{i} log\frac{P_{i}}{Q{i}}
\]
where $P_{i}$ and $Q_{i}$ are probabilities of outcome $i$ for a reference and test distributions respectively. In context of MSMs, the relative entropy between a reference and test MSM can be defined as:
\begin{equation}
    \mathcal{D}(P\vert \vert Q) = \sum \limits_{i,j}^{N} P_{i} P_{ij}  log\frac{P_{ij}}{Q_{ij}}
    \label{eq3}
\end{equation}
where $P$ and $Q$ are transition probability matrices of the reference and test MSM respectively. And $P_{i}$ is the stationary probability of state $i$.  A lower value of $\mathcal{D}$ would indicate similarity to the reference transition matrix. In our case, reference MSM is the converged ground truth MSM and the test MSM is the MSM constructed at each round of adaptive sampling for each policy respectively. To construct the test MSM, the state definition from the ground truth MSM is enforced to keep the order of $P$ and $Q$ the same. In addition, a uniform prior $1/N$, where N is the number of states, in the form of pseudo-counts is added to the count matrix of the test MSM before row normalization. The purpose of the uniform prior is two fold. Firstly, it prevents the relatively entropy metric in (\ref{eq3})
from becoming undefined when a zero value in encountered. In addition, at least a single count must be observed for any particular state, but before sampling the probability of this count to transition to another state is uniform, hence the choice of the prior.

In order to combine these two metrics, we define the total loss function as follows:
\[
    \mathcal{L} = \beta (1 - \mathcal{E}) + (1-\beta)\mathcal{D}
\]
or more explicitly
\begin{equation}
    \mathcal{L} = \beta \left(1 - \frac{S_{v}}{S_{T}}\right) + (1-\beta) \sum \limits_{i,j}^{N} P_{i} P_{ij}  log\frac{P_{ij}}{Q_{ij}}
    \label{eq4}
\end{equation}
where $\beta$ is the parameter controlling the contribution of the exploration loss $(1-\mathcal{E})$ and convergence loss $\mathcal{D}$ to the total loss $\mathcal{L}$.

The most \textit{optimal} policy is thus identified using:
\[
p'=\arg\min\limits_{p \in \mathcal{P}} \mathcal{L}(p)
\]
Consequently, sampled data from $p'$ is appended to the previous round data and the cycle is iterated. This workflow is summarized in Algorithm \ref{alg1}.

\begin{algorithm}[!h]
\caption{Optimal Sampling via Policy Ranking}\zlabel{methods-1}
\begin{algorithmic}[1]
\REQUIRE{
    potential $V(\boldsymbol{x})$, 
    clustering model $\mathcal{C}$,
    policy ensemble $\mathcal{P}$,
    loss function $\mathcal{L}$,
    number of epochs $E$,
    trajectory length $T$,
    trajectories per epoch $M$
}    
\STATE Sample initial data $X^0$ starting from $V(\boldsymbol{x}_0)$
\FOR{$e$ in $1 \dots E$}
\STATE $D^{e-1} = \mathcal{C}(X^{e-1}) = [\mathcal{C}(\boldsymbol{x}_0) \dots \mathcal{C}(\boldsymbol{x}_{eTM})] = [\boldsymbol{d}_0 \dots \boldsymbol{d}_{eTM}]$
    \FOR{$p$ in $\mathcal{P}$}
    \STATE $\boldsymbol{x}_{p} = p(D^{e-1})$
    \STATE Sample new data ,$X_{p}$  starting from $V(\boldsymbol{x}_p)$
    \STATE $\boldsymbol{l}_{p}$ = $\mathcal{L}(X_{p})$
    \ENDFOR

\STATE$ p'=\arg\min\limits_{p \in \mathcal{P}} {L}(p)$, where $L=\{\boldsymbol{l}_{p}: \forall p \in \mathcal{P}\}$
\STATE Concatenate $X_{p'}$ to $X^{e-1}$ to obtain $X^e$
\ENDFOR
\end{algorithmic}
\label{alg1}
\end{algorithm}

\section{Results and Discussion}\zlabel{results}

\subsection{Choosing a different policy at each round outperforms single policy sampling}

\begin{figure}[!htbp]
    \centering
    \includegraphics[width=0.8\textwidth]{ 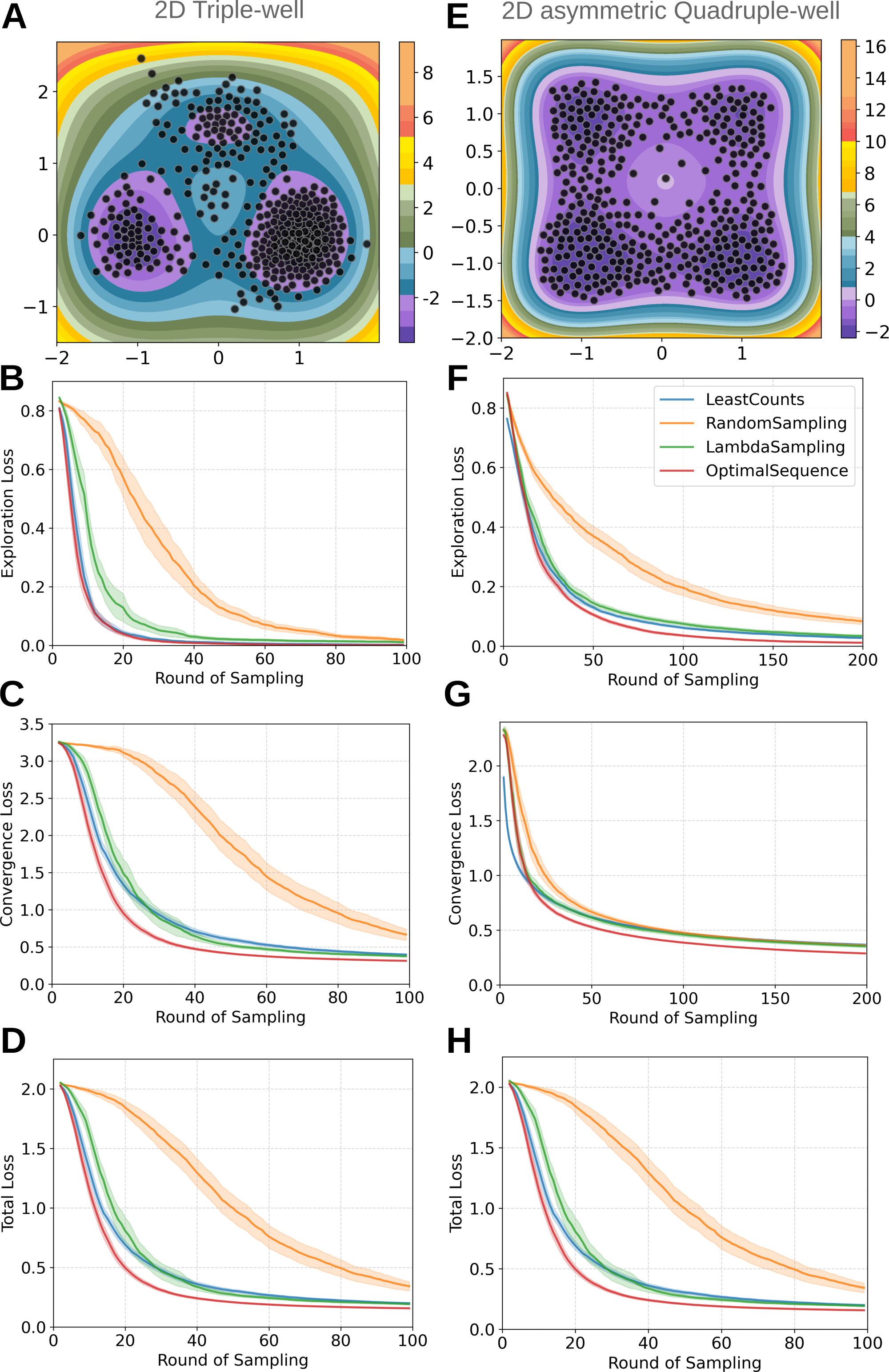}
    \caption{Comparison of policy ranking with single policy selection on toy potentials. (A) 2D triple-well potential. (E) Asymmetric quadruple-well potential with four minima, all with different potential depth. Black dots represent clusters from the ground truth sampling. (B)(C)(D) represent exploration loss, convergence loss and total loss respectively as in eq(\ref{eq4}) for triple-well system.(F)(G)(H) represent exploration loss, convergence loss and total loss respectively for the asymmetric quadruple-well system. Shaded regions indicate 95\% CI. }
    \label{fig:figure_3}
\end{figure}

To investigate the existence of a set of policy sequences which would outperform single policy sampling, we chose two toy systems. Figure \ref{fig:figure_3}A showcases a \texttt{2D triple-well} potential. The system simulates particle dynamics in a 2D coordinate system. The governing stochastic differential equation is:
\[
d\mathbf{X}_t = \nabla V(\mathbf{X}_t) \, dt + \sigma(t, \mathbf{X}_t) \, d\mathbf{W}_t 
\]

where \(\mathbf{W}_t\) is a Wiener process, stochastic force parameter is set to \(\sigma = 1.09\), and the potential \(V\) is:
\[
V(\mathbf{x}) = 3e^{-x^2 - \left(y - \frac{1}{3}\right)^2} - 3e^{-x^2 - \left(y - \frac{5}{3}\right)^2} 
- 5e^{-\left(x - 1\right)^2 - y^2} - 5e^{-\left(x + 1\right)^2 - y^2} 
+ \frac{2}{10}x^4 + \frac{2}{10}\left(y - \frac{1}{3}\right)^4.
\]
The integration time-step was kept at $h=1e-5$ and number of steps between each evaluation were set to $n\_steps=50$. Length of each generated trajectory was fixed at $300.$ 

The second system we studied was a \texttt{asymmetric quadruple-well} potential (Figure \ref{fig:figure_3}E). With the potential \(V\) being given by
\[
V(\mathbf{x}) = \left( x_1^4 - \frac{16}{3}x_1^3 - 2x_1^2 + \frac{3}{16}x_1 \right) + \left( x_2^4 - \frac{8}{3}x_2^3 - 2x_2^2 + \frac{3}{8}x_2 \right).
\]
The stochastic force parameter was set to \(\sigma = 0.6\). The integration time-step was kept at $h=1e-3$ and number of steps between each evaluation was set to $n\_steps=200$. Length of each generated trajectory was fixed at $50.$ 

Sampling was performed using the optimal sampling algorithm described in Section \ref{sec:methods}. To compare with single policies, we also performed adaptive sampling for individual policies in $\mathcal{P}$. For all cases  each round of adaptive sampling consisted of 5 parallel trajectories. 

For \texttt{2D triple-well} potential, a total of 30 replicates were performed, where each replicate consisted of 100 rounds of adaptive sampling. While for the \texttt{asymmetric quadruple-well} potential, 50 replicates of 100 rounds each were carried out.

Figures \ref{fig:figure_3}A and E showcase the potential landscapes for both toy systems. To compute relative entropy via (\ref{eq3}), a reference MSM was needed as a ground truth for comparison. The black dots represent the resulting clusters of this reference MSM, the cluster number was fixed at 400 for both systems. The implied time-scale convergence and Chapman-Kolmogorov test results are illustrated in SI Figures.

Figures \ref{fig:figure_3}B , C and D showcase the performance of the policy ranking scheme described earlier, compared to single policy sampling for the triple-well potential. Figure \ref{fig:figure_3}B shows, for the exploration loss, that the Optimal Sequence sampling, i.e. the sampling derived from the policy ranking scheme, closely follows the Least Counts sampling which outperforms other policies. For convergence loss, which determines how similar is the MSM from the current sampling to the ground truth MSM, Figure \ref{fig:figure_3}C shows relatively similar performance for both Least Counts and Lambda Sampling. Whereas policy ranking here outperforms the best performing policies consistently as the rounds of sampling increase. Figure \ref{fig:figure_3}D shows the combined effect of the exploration and convergence losses with $\beta=0.5$ in (\ref{eq4}) for the optimal sequence sampling. Figure \ref{fig:figure_3}F, G and H showcase the performance of the policy ranking scheme described earlier, compared to single policy sampling for the quadruple-well potential. For this test case, policy ranking consistently gives faster convergence and exploration compared to the single policies. Compared with the triple-well potential where policy ranking showed an identical performance to Least Counts for exploration, in this case policy ranking outperforms Least Counts, especially towards the latter sampling stage where Least Counts fails to sample the rare states.  

\subsection{Ranking policies on physical systems}
\begin{figure}[!ht]
    \centering
    \includegraphics[width=\textwidth]{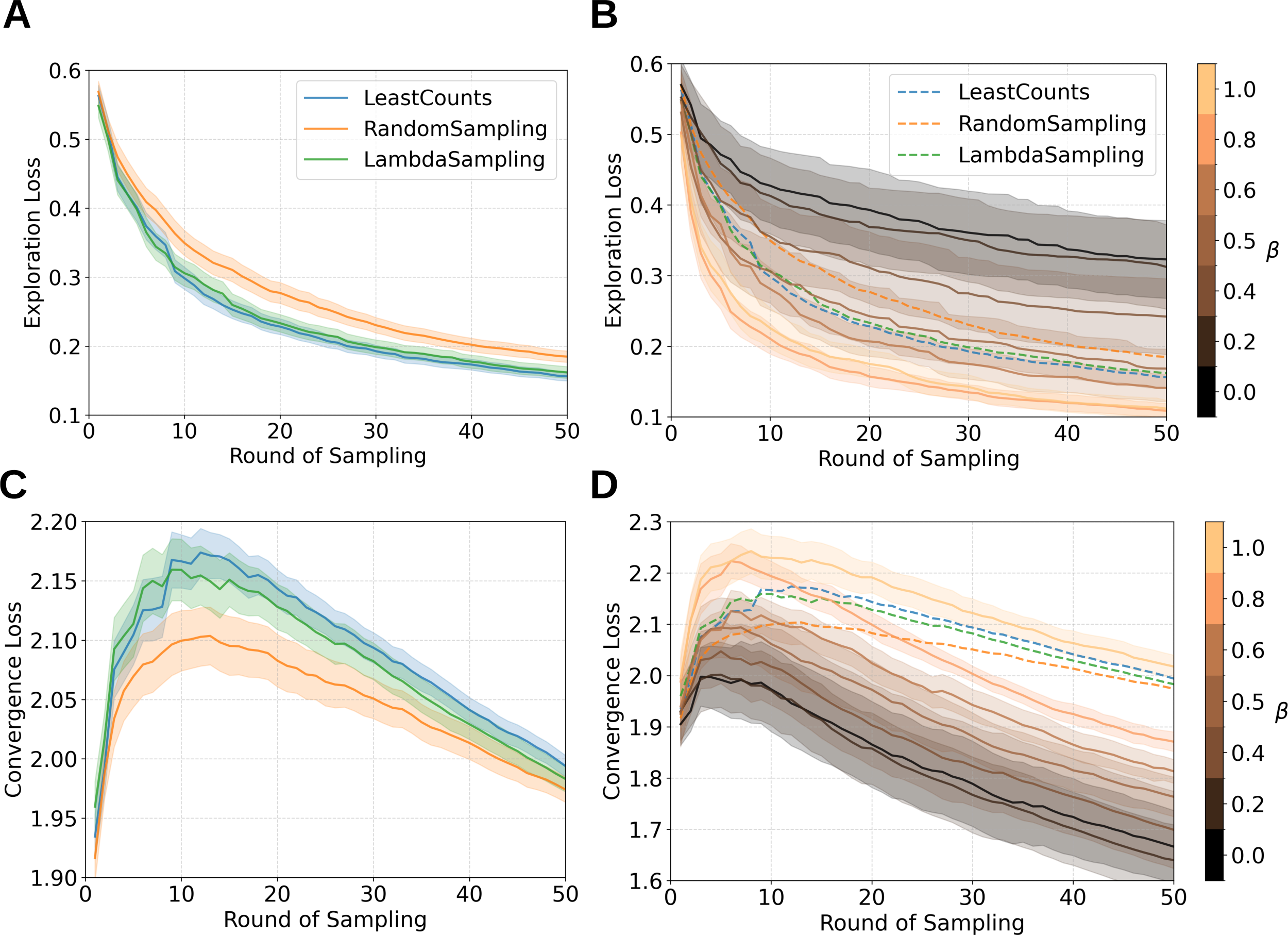}
    \caption{Comparing policy ranking to single policy performance for alanine-dipeptide.(A)(C) Exploration and convergence losses for single policies. (B)(D) Single policy losses are shown in dotted lines. While optimal sequence losses are shown in sequential colormap with varying values of $\beta$ in (\ref{eq4}) to weigh relative importance of exploration and convergence of the ranking scheme. Shaded regions indicate 95\% CI.}
    \label{fig:figure_4}
\end{figure}
\subsubsection{Alanine-dipeptide}
Moving on towards more realistic test systems, we compare the policy ranking algorithm with single policies on alanine-dipeptide system. For details on the ground truth model as well as simulation details, refer to the SI methods. In contrast to toy potentials, where the 2D trajectories were considered realizations of a 2D particle in those potentials. In case of alanine-dipeptide, the backbone dihedrals $\phi$ and $\psi$  were chosen as reduced representations (collective variables). Subsequently, clustering and seed selection was carried out in the dihedral space.  The number of clusters was fixed at 300.

For each policy, 50 rounds of adaptive sampling were carried out. With each round consisting of 5 short trajectories of 4ps each. For the single policies 120 replicates of 50 rounds each were performed. In case of policy ranking scheme however, instead of fixing $\beta=0.5$, we carried out 20 replicates of 50 rounds each for a range of $\beta$ values between $(0,1)$ inclusive. The idea here was to investigate if selection of the best policy according to an asymmetric contribution of exploration and convergence would differ from the case where $\beta=0.5$.

 Figure \ref{fig:figure_4}A shows the exploration loss for the single policies. As previously observed from the toy system performance, Random Sampling again gives the poorest performance in exploration compared to Least Counts and Lambda Sampling which perform within error range. Figure \ref{fig:figure_4}B compares this single policy performance (shown in dashed lines) to optimal sequence from policy ranking at different values of $\beta$. From (\ref{eq4}) when $\beta \rightarrow 0$, the contribution of exploration loss diminishes ($\mathcal{E} \rightarrow 0$ ) and convergence loss becomes dominant ($\mathcal{L} \rightarrow \mathcal{D}$). Conversely when $\beta \rightarrow 1$ the convergence loss disappears and best exploratory policy at a given round is chosen. In  Figure \ref{fig:figure_4}, the policy ranking performance is as expected with varying values of $\beta$. Higher values of $\beta$ consistently improve exploration performance. In comparison with single policies, optimal sequences with $\beta > 0.5$ outperform the best performing single policies, whereas lower values of $\beta$ perform poorly and with significantly higher variance.

  Figures \ref{fig:figure_4}C and D showcase the convergence losses for single policies and policy ranking respectively. For single policy ranking in  Figure \ref{fig:figure_4}C, Random sampling provides the fastest convergence in the first half of sampling, and rate of convergence seems to converge as adaptive sampling rounds are increased.  Figure \ref{fig:figure_4}D shows the comparison of policy ranking performance in terms of convergence loss, with respect to single policies (showed in dashed lines). Performance trends are inverted for convergence loss as higher values of $\beta$ will favor exploration and vice versa for convergence. It is interesting to note here however that all values of $\beta$ in (0,0.8) outperform single policy sampling and only $\beta=1$ shows a poorer performance. This shows that multiple values of $\beta$ can deliver improved sampling in terms of exploration and convergence both, compared to single policy sampling.

Now that we understand that an optimal sequence of policies chosen via a ranking scheme outperforms single policies. We are interested to see how the total loss function (equation (\ref{eq4})) affects policy selection at different values of $\beta$. Using the 20 replicates of simulations for optimal sequence at different $\beta$ values, we record the occurrences of different policies at each round.  Figure S10 shows the probability distribution of these occurrences. The color intensity of each cell illustrates the probability of a particular policy being selected at that round. For higher values of $\beta$, the total loss function prioritizes policies performing better in exploration and vice versa for convergence. However the heat-map does not show a single policy being selected throughout for any values of $\beta$. For $\beta=1$, the results show a higher tendency of the algorithm to select Least Counts, which is the policy showing the best performance amongst the single policies (Figure \ref{fig:figure_4}A), however even this choice is not consistent and there is a finite chance of other policies being selected, which actually leads the optimal sequence to outperform Least Counts. Conversely, for smaller values of $\beta$, Figure \ref{fig:figure_4}C showed Random Sampling to give a faster convergence, but Figure S10 shows that the algorithm distributes the policy choice between the three policies to achieve convergence trends which are significantly faster than Random Sampling. These insights offer an explanation for the question we started out with, i.e. do there exist sequences of policy choices that deliver better performance then single policy choice at each round? We find that not only do these sequences exist, but the set of such sequences is diverse and utilizes all policies in the ensemble.

\subsubsection{Sampling Activation related dynamics of Class F GPCR}
\begin{figure}[!ht]
    \centering
    \includegraphics[width=\textwidth]{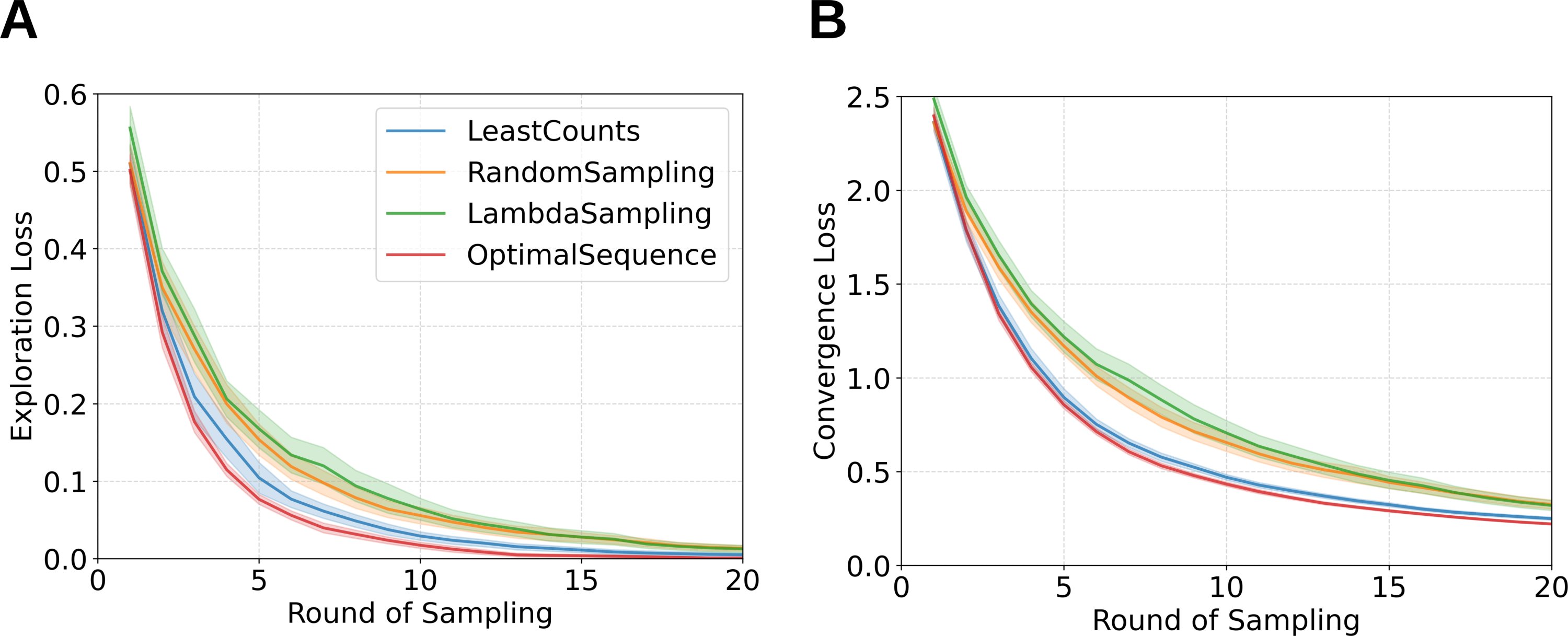}
    \caption{Comparing policy ranking to single policy performance for Smoothened (GPCR).(A) Exploration loss of single policies compared to optimal sequence sampling (red). (B) Convergence loss single policies compared to optimal sequence sampling (red) .Shaded regions indicate 95\% CI.}
    \label{fig:figure_5}
\end{figure}

We have shown that for 2D toy potentials as well as alanine-dipeptide, optimizing policy choice at each round outperforms single policy sampling in each case. To demonstrate the idea further on a realistic biomolecular system, we sample the activation pathway for a G-Protein Couples receptor, Smoothened (SMO). Smoothened is a member of the Frizzled family of G-Protein Coupled Receptors(GPCRs). SMO plays an important role in the Hedgehog signaling pathway. This pathway is fundamental for preserving stability throughout critical biological functions such as cell differentiation, adult tissue regeneration, and the development of embryos \cite{Briscoe2013,Lee2016}. Markov state models along with adaptive sampling schemes extensively to investigate the GPCR dynamics\cite{Deganutti2024, Bansal2023, Dutta2023, Song2021, Dutta2022, Dutta2022a, Kohlhoff2013}. Therefore, we have chosen a recent study by Kim \textit{et al.} (2024)\cite{Kim2024} on the activation of the class F GPCR, Smoothened bound to the ligand, Cyclopamine. This study showed that cyclopamine acts as an agonist when bound to the Cysteine Rich Domain (CRD) of the Smoothened receptor and as an antagonist when bound to the Transmembrane Domain (TMD). Simulations revealed that CRD binding promotes receptor activation, TMD binding induces antagonism, and simultaneous binding to both sites results in a weak antagonistic effect with intermediate behavior. Here, we use the Markov State Model reported in this study for the case where cyclopamine is bound to TMD. Using this MSM we generated trajectories using kinetic Monte Carlo (kMC) sampling. This scheme performs stochastic propagation of the system according to the kinetic information in the MSM. The system was represented with 150 states. Please refer to SI methods for details of MSM construction. For the kMC scheme, 30 replicates of 20 adaptive sampling rounds each were performed. Deeptime\cite{hoffmann2021deeptime} library was used for the kMC propagation. Each adaptive sampling round consisted of 5 trajectories of 80 steps each. For each round, a random starting state was chosen to get the initial sampling.

Figure \ref{fig:figure_5} shows the performance of optimal sequences derived from policy ranking compared to single policy performance. For both exploration and convergence losses, Least Counts outperforms Lambda and Random Sampling, this is expected because using kinetic Monte Carlo sampling, the simulations represent transition between predefined discrete states, like a graph topology. Optimal sequences of policies derived from ranking however outperform Least Counts in both metrics as clearly evidenced by the plots.

\subsection{Ranking policies on the fly}
Our results have illustrated that on diverse systems, from 2D potentials to GPCRs, that a different policy selection at each round significantly outperforms single policy sampling. Algorithm \ref{alg1} and Figure \ref{fig:figure_2} illustrate the methodology that was followed. This scheme provides an idea of an ensemble view of policy space, instead of a single \textit{best} policy for adaptive sampling. Despite proving that an optimal sequence of policies exists, this technique falls short of identifying that sequence or set of sequences on the fly. Such a technique would have to identify (or approximate the performance of) this set of sequences without the need of running simulations from all available policies in $\mathcal{P}$ at each round.

As discussed earlier, the policy space with \textit{p} policies and \textit{n} sampling rounds evolves exponentially as $p^n$. The task is computationally hard given the size of the policy space. In addition, Figure S10 demonstrated that the algorithm does not focus on a single policy(or even a single sequence of policies) during sampling of a particular region of the landscape or a particular round of sampling, therefore it is a challenging task to identify a single sequence by a \textit{learning} technique. In this section, we propose two techniques that aim to approximate these sequences on the fly.

\subsubsection{Random walk in policy space}
Given the findings from policy ranking, it is obvious that different policy selection improves the performance for adaptive sampling. A random walk in policy space would therefore be a strong baseline. The idea is to perform random sampling from $\mathcal{P}$ to select the policy for each  round of adaptive sampling. We term this strategy Random Policy Sampling.

\subsubsection{Ensemble Adaptive Sampling Scheme}
\begin{figure}[!ht]
    \centering
    \includegraphics[width=\textwidth]{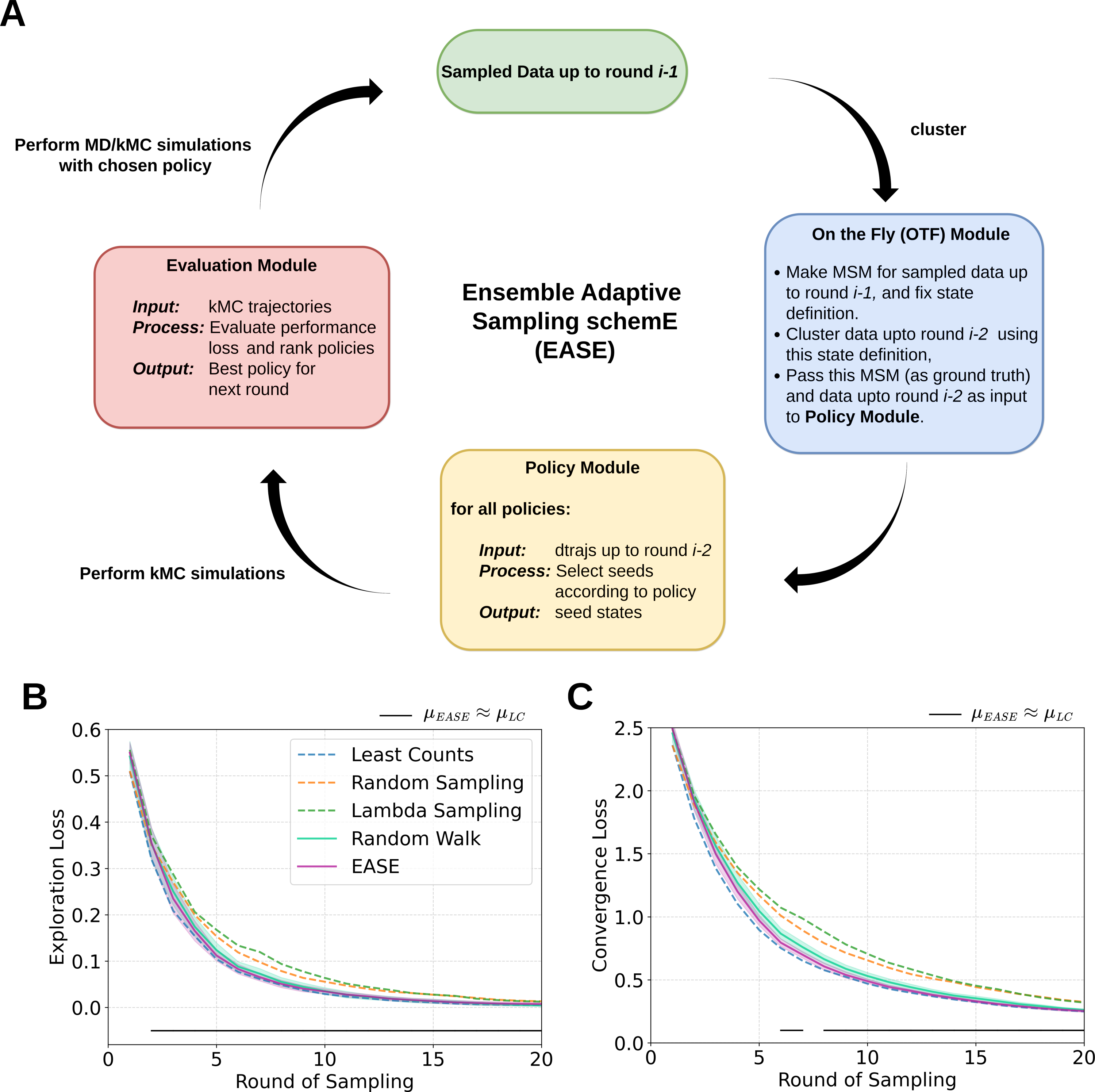}
    \caption{(A) Workflow of Ensemble Adaptive Sampling Scheme (EASE). Sampled data from previous rounds is used to perform on-the-fly ranking by performing kinetic Monte Carlo simulations. (B) Exploration loss for EASE and random walk sampling compared to single policies (C) Convergence loss for EAST and random walk sampling compared to single policies. Shaded regions indicate 95\% CI and horizontal black marker indicates rounds where performance of EASE is statistically same as the best performing single policy, i.e. Least Counts.}
    \label{fig:figure_6}
\end{figure}

In order to rank policies on the fly, we employ a scheme that mimics a non-linear auto-regressive model. The scheme uses exploration and convergence performance to guide the search of an optimal policy in the policy ensemble space, termed EASE (Ensemble Adaptive Sampling schemE). The idea here is to approximate the \textit{optimal} policy at \texttt{round(i)} using sampling from  \texttt{round (i-1)} and \texttt{round(i-2)}. The workflow is illustrated in Figure \ref{fig:figure_6}A. At \texttt{round(i-1)}, state clustering is performed on accumulated data till \texttt{round (i-1)}. In the On the Fly (OTF), module a markov state model is constructed using this state definition. Accumulated data up to \texttt{round(i-2)} and this MSM are passed on to the \texttt{Policy Module}. As before, the policy module outputs seed states according to different policies in the policy ensemble $\mathcal{P}$. In contrast with the policy ranking scheme, kMC simulations are performed from these seed states, using the MSM generated in the OTF module. The idea is that kMC simulations offer a substantial reduction in wall-time, thereby allowing to perform simulations for each policy in ensemble $\mathcal{P}$. These trajectories are then ranked according to (\ref{eq4}) and the most \textit{optimal} policy is selected to perform \texttt{round(i)} simulations. Consequently the process is repeated till required sampling is achieved.     

We test  Random Policy Sampling and EASE on Smoothened receptor activation process. Figure \ref{fig:figure_6} B,C show the exploration and convergence losses for these two techniques. The single policies performance is reproduced here and represented by dashed lines. It is interesting to observe that Random Policy Sampling (shown in teal), which represents a random walk in policy space outperforms Lambda and Random sampling policies for most of the 20 rounds. This reaffirms our conclusions from previous results that without \textit{a priori} knowledge of a single best policy, an ensemble approach gives better performance. EASE also outperforms Lambda and Random Sampling but also reaches on par performance as Least Counts, which was the best single policy for sampling of this system. This performance can be explained by the probability distributions of the policies selected by these two schemes. Figures S7, S8 and S9 show the probability distribution for policy ranking scheme, random walk scheme and EASE, respectively. It is evident that EASE captures the overall distribution of the policy selection from policy ranking scheme, focusing more on Least Counts, whereas Random Policy sampling, by definition, focuses on all policies relatively evenly. 

\section{Conclusions}\zlabel{conclusions}
In this study, we propose an ensemble approach to adaptive sampling for biomolecular simulations. We develop the policy ranking algorithm that selects a policy sequence by ranking policies at each round of sampling. Our results show that the choice of a different policy at each round significantly outperform single policies. Analysis of selection probabilities illustrates that a combination of sub-performing policies performs better for both exploration and convergence metrics, and we also observe that a set of this sequence of policies exist which the ranking scheme identifies instead of fixating on a single sequence.

Given the enormity of the policy space that scales exponentially with the number of adaptive sampling rounds, it is a challenging task to approximate the set of optimal policy sequences without performing MD simulations for the whole policy ensemble. We proposed two schemes for on-the-fly ranking of policies. A  random walk in policy space, that selects a policy randomly and EASE (Ensemble Adaptive Sampling schemE), that approximates the policy ranking scheme by performing kinetic Monte Carlo simulations. EASE identifies the most \textit{optimal} policy at the last round, and uses it for sampling at the current round. We observe that without any \textit{a priori} ranking, EASE is able to replicate performance of Least Counts which delivers the best performance, whereas random walk scheme outperforms the other single policies. 

Although EASE identifies and performs on-par with the best performing single policy, but it does not deliver the same performance which the policy ranking scheme promises. EASE also employs kinetic Monte Carlo (kMC) simulations as a ranking scheme. We anticipate that Least Counts will generally outperform other policies in graph-mimicking structures like trajectories generated from Markov State Models, therefore EASE will demonstrate a tendency towards achieving better exploration but subpar convergence on systems where Least Counts is not the best performing single strategy in both metrics.

The policy ranking framework can also be improved upon and modified in future work. The policy ensemble $\mathcal{P}$ can be extended to include diverse policies, such as reinforcement learning inspired schemes. The released code is modular and allows for easy implementation of custom adaptive sampling policies. The total loss function presented in eq\ref{eq4} can be modified by scaling the exploration and convergence losses or by introducing other functional forms. This opens exciting new areas of research in this field, where Reinforcement Learning inspired methods could better approximate the policy ranking scheme and deliver improved performance on the fly.

\begin{acknowledgement}
D.S. acknowledges support from the Army Research Office under Cooperative Agreement No. W911NF-22-2-0246. The views and conclusions contained in this document are those of the authors and should not be interpreted as representing the official policies, either expressed or implied, of the Army Research Office or the U.S. Government. The U.S. Government is authorized to reproduce and distribute reprints for Government purposes notwithstanding any copyright notation herein. D.S. also acknowledges funding from the National Institutes of Health MIRA Award R35GM142745.
\end{acknowledgement}

\begin{suppinfo}

The Supporting Information is available free of charge at \url{http://pubs.acs.org}.
Code necessary to reproduce the simulations is available on \url{https://github.com/ShuklaGroup/EASE}.
\end{suppinfo}

\bibliography{acs-achemso}

\end{document}